\documentclass[twocolumn]{emulateapj}
\usepackage{lscape}

\shorttitle{Star-Formation and Size at $\mathbf z \sim 2.5$}
\shortauthors{Zirm et al.}

\def\spose#1{\hbox to 0pt{#1\hss}}
\def\simlt{\mathrel{\spose{\lower 3pt\hbox{$\mathchar"218$}}
     \raise 2.0pt\hbox{$\mathchar"13C$}}}
\def\simgt{\mathrel{\spose{\lower 3pt\hbox{$\mathchar"218$}}
     \raise 2.0pt\hbox{$\mathchar"13E$}}}

\journalinfo{Accepted for publication in the Astrophysical Journal, 6
November 2006}

\begin{document}

\title{NICMOS Imaging of DRGs in the HDF-S: A Relation Between
Star-Formation and Size at $z \sim 2.5$\altaffilmark{1}}

\author{Andrew W. Zirm\altaffilmark{2,3}, A.~van der
Wel\altaffilmark{4}, M.~Franx\altaffilmark{2},
I.~Labb{\'e}\altaffilmark{5,6}.  I.~Trujillo\altaffilmark{7}, P.~van
Dokkum\altaffilmark{8}, S.~Toft\altaffilmark{9},
E.~Daddi\altaffilmark{10}, G.~Rudnick\altaffilmark{11,12},
H.-W.~Rix\altaffilmark{13}, H.J.A.~R{\"o}ttgering\altaffilmark{2},
P.~van der Werf\altaffilmark{2}}

\altaffiltext{1}{Based on observations with the NASA/ESA Hubble Space
Telescope, obtained at the Space Telescope Science Institute, which is
operated by the Association of Universities for Research in Astronomy,
Inc., under NASA contract NAS 5-26555. These observations are
associated with program \# 9723.}

\altaffiltext{2}{Leiden Observatory, Leiden University, Postbus 9513,
NL-2300 RA Leiden, The Netherlands}

\altaffiltext{3}{Current address: Johns Hopkins University, 3400
N. Charles Street, Baltimore, MD, 21218; {\tt azirm@pha.jhu.edu}}

\altaffiltext{4}{Johns Hopkins University, 3400 N. Charles Street,
Baltimore, MD, 21218}

\altaffiltext{5}{Carnegie Observatories, 813 Santa Barbara Street,
Pasadena, CA 91101} 

\altaffiltext{6}{Carnegie Fellow}

\altaffiltext{7}{School of Physics \& Astronomy, University of
Nottingham, University Park, Nottingham NG7 2RD, UK}

\altaffiltext{8}{Department of Astronomy, Yale University, New Haven,
CT 06520-8101} 

\altaffiltext{9}{European Southern Observatory,
Karl-Schwarzschild-Strasse 2, D-85748 Garching, Germany}

\altaffiltext{10}{CEA/Saclay, Service d'Astrophysique, Orme des
Merisiers, 91191 Gif-sur-Yvette, France}

\altaffiltext{11}{National Optical Astronomy Observatory, 950 North
Cherry Avenue, Tucson, AZ 85719}

\altaffiltext{12}{Goldberg Fellow}

\altaffiltext{13}{Max-Planck Institute for Astronomy, K{\"o}nigstuhl
17, 69117 Heidelberg, Germany}

\begin{abstract}
  We present deep, high angular-resolution {\it HST}/NICMOS imaging in
  the Hubble Deep Field South (HDF-S), focusing on a subset of 14
  Distant Red Galaxies (DRGs) at $z \sim 2.5$ galaxies that have been
  pre-selected to have $J-K > 2.3$.  We find a clear trend between the
  rest-frame optical sizes of these sources and their
  luminosity-weighted stellar ages as inferred from their broad-band
  spectral energy distributions (SEDs).  Galaxies whose SEDs are
  consistent with being dusty and actively star forming generally show
  extended morphologies in the NICMOS images ($r_{e} \simgt 2$ kpc),
  while the 5 sources which are not vigorously forming stars are
  extremely compact ($r_{e} \simlt 1$ kpc).  This trend suggests a
  direct link between the mean ages of the stars and the size and
  density of the galaxies and supports the conjecture that early
  events quench star-formation and leave compact remnants.
  Furthermore, the compact galaxies have stellar surface mass
  densities which exceed those of local galaxies by more than an order
  of magnitude.  The existence of such massive dense galaxies presents
  a problem for models of early-type galaxy formation and evolution.
  Larger samples of DRGs and higher spatial resolution imaging will
  allow us to determine the universality of the results presented here
  for a small sample.
\end{abstract}

\keywords{galaxies: high-redshift --- galaxies: fundamental parameters
--- galaxies: evolution --- galaxies: formation --- infrared:
galaxies}

\section{Introduction}

In the local universe, the star-formation rate per stellar mass
(specific star-formation rate) correlates strongly with galaxy
concentration and stellar mass surface density
\citep*[$\sigma_{50}$;][]{Kauffmannetal03b,Brinchmannetal04}.  The
high concentration, high $\sigma_{50}$ early-type galaxies may result
from the relationship between surface gas density and star-formation
rate (or efficiency) during their formation epoch.  In any case, these
relations are likely indicative of a fundamental principle of galaxy
formation and evolution.  By determining whether these, or similar,
relations hold during earlier cosmic epochs we may address questions
such as: Where are the majority of stars formed?  Is the same event
which truncates star-formation in early-type galaxies responsible for
their morphological transformation, i.e., are there red disk galaxies
and/or blue early-types?

A vast majority of the known UV-selected high redshift ($z \simgt 2$)
galaxies are rapidly forming stars.  It is perhaps not surprising
then, that most of these galaxies have irregular or spiral
morphologies \citep*{GiavaliscoSteidelMacchetto96,LabbeetalDisk03,
Chapmanetal03,Conseliceetal04,Papovichetal05}.  This is only half of
the story.  To test whether a morphology--star-formation relation
exists at high redshift one must first identify the more quiescent
galaxies at similar redshifts, if they exist.  Searches for passive
galaxies benefit by moving to the rest-frame optical where the light
is not dominated by young stars.  At high-redshift this requires
observations in the near-infrared or longer wavelengths.  Only
recently have deep near-infrared (rest-frame optical) studies been
undertaken using 8m-class telescopes.  The Faint Infrared
Extragalactic Survey (FIRES; Franx et al. 2003\nocite{Franxetal03})
using ESO/VLT, for example, has discovered a class of galaxies
(Distant Red Galaxies; DRGs) which are not as UV bright as previously
optically-selected galaxies at $z \sim 2-3$.  These sources are
selected using the color-cut $(J-K)_{\rm Vega} > 2.3$ which
corresponds to rest-frame $U-V$ at $z=2.5$.  Galaxies can have red
$U-V$ colors for primarily two reasons: they can have luminous
starbursts which are highly dust reddened or they can have a large
mass of evolved stars \citep*{Cimattietal02, Franxetal03,
vanDokkumetal04, Labbeetal05}.  The brightest of these DRGs have now
been confirmed and studied spectroscopically in the near-infrared
\citep*{Krieketal06a, Krieketal06b}.

By combining rest-frame optical sizes and multi-wavelength spectral
energy distributions we can begin to discern the evolutionary pathways
of galaxies at various redshifts
\citep*[e.g.,][]{GEMS,Nachoetal04,Nachoetal06}.  At redshifts greater
than unity, this requires excellent data in the near-infrared for both
object selection and morphologies.  Previous studies of NIR-selected
galaxy sizes have derived general evolutionary trends using
ground-based (lower spatial resolution) data and have not addressed
the relation between galaxy size and star-formation
\citep*[e.g.,][]{Nachoetal04,Nachoetal06}.  In this article we present
deep, high spatial-resolution near-infrared imaging of 14 of these
DRGs using the Near-infrared Camera and Multi-object Spectrograph
(NICMOS) on-board the {\it Hubble Space Telescope} ({\it HST}).  These
images cover the HDF-S \citep*[][]{Williamsetal00,Casertanoetal00} in
the F160W ($H_{160}$) passband and are of sufficient quality to
determine the rest-frame optical sizes of this recently discovered
population of galaxies.  The availability of broad-band imaging data
out to rest-frame $K$-band allows us to relate galaxy size to the SED
We use a cosmology with $(\Omega_{\rm m},\Omega_{\Lambda}) =
(0.27,0.73)$ and $H_0 = 71$ ${\rm km s^{-1} Mpc^{-1}}$ throughout.  At
$z = 2.5$ one arcsecond subtends 8.2 physical kpc and one NICMOS
Camera 3 resolution element (FWHM $\sim 0\farcs26$) corresponds to 2.1
kpc.  The stellar masses inferred via spectral fitting are derived
using a Salpeter initial mass function with mass range $0.1-100
M_{\odot}$.

\section{NICMOS Imaging and Sample Selection}

The primary dataset used in this paper is NICMOS imaging of the Hubble
Deep Field South (HDF-S).  Eight pointings of NICMOS Camera 3 were
required to cover the full WFPC2 field of the HDF-S.  Camera 3 has a
field-of-view of approximately 50$\arcsec$ on a side at a pixel scale
of $0\farcs2$ ${\rm pixel^{-1}}$.  Each of these 8 pointings was
imaged using a six point, sub-pixel dither pattern to better sample
the point-spread function (PSF).  The individual exposures were
reduced in the usual fashion using the pipeline within IRAF and taking
particular care to mask out deviant pixels.  The dither offsets were
determined using cross-correlation and inter-pointing offsets were
measured using the ground-based VLT/ISAAC imaging data as a reference
frame.  These shifts were used as input to the {\it drizzle} task in
IRAF which was used to create the full mosaic of the field.  The
combined data reach a 3$\sigma$ depth of 25.0 AB magnitudes in a
$0\farcs5$ radius circular aperture and have an average integration
time of 5200 s.

We focus the current study on the 14 Distant Red Galaxies (DRGs) in
this field.  The results presented here depend on previous work,
primarily rest-frame near-infrared imaging from {\it Spitzer}/IRAC and
the subsequent broad-band spectral energy distribution (SED) fits
\citep*{Labbeetal05}.  We follow the object numbering used by
Labb{\'e} et al. (2005).  These broad-band data have been used to
determine the photometric redshifts, stellar masses and star-formation
properties of these DRGs
\citep*{Rudnicketal01,Rudnicketal03,Labbeetal05,Wuytsetal06}.  We note
that we use the Rudnick et al. (2003) photometric redshifts rather
than the revised versions used in Rudnick et
al. (2006)\nocite{Rudnicketal06} which are systematically higher.
Object 327 is blended with a nearby source in the $K_{S}$-band data.
We present the NICMOS data and profile fits for 327 but exclude it
from any analysis dependent on the longer wavelength data.  Objects 66
and 810 had relatively poor SED fits.  For 66 this is likely due to
emission-line contamination of the broad-band fluxes, a conjecture
confirmed by spectroscopy.  However we expect that its derived stellar
mass is relatively unaffected.  For 810 the source of the poor fit is
less clear and we therefore qualify its high mass and lack of
star-formation as tentative claims.

\section{Galaxy Sizes and Morphologies}

We have determined galaxy sizes by fitting PSF-convolved, analytic
surface-brightness profiles to each of the 14 DRGs in the HDF-S
sample.  For each galaxy we use individually generated TinyTim
\citep*{TinyTim} model point-spread functions (PSFs).  The final PSFs
were combined in the same manner as the data itself to account for
both the variation of the instrumental response over the detector and
the dependence of the reconstructed PSF on the {\it drizzle}
algorithm.  To assess the dependence of our results on the assumed PSF
we have also constructed a PSF from a suitably bright and isolated
star in our final mosaic.  We then fit full two-dimensional convolved
models to the data using both an ``in-house'' code \citep*{Franx1993,
vanDokkumFranx96} and the publicly-available GALFIT code
\citep*{GALFIT} with both the model and stellar PSFs.  By fitting both
model PSFs and a star to other stars in the field we derived a
(conservative) minimum measurable size of approximately $0\farcs06$ or
half one pixel.  All of the derived galaxy sizes are above this limit.
We used the data pixel weights to optimize the fits and the SExtractor
object segmentation map to mask neighboring sources.

\begin{figure}[t]
\plottwo{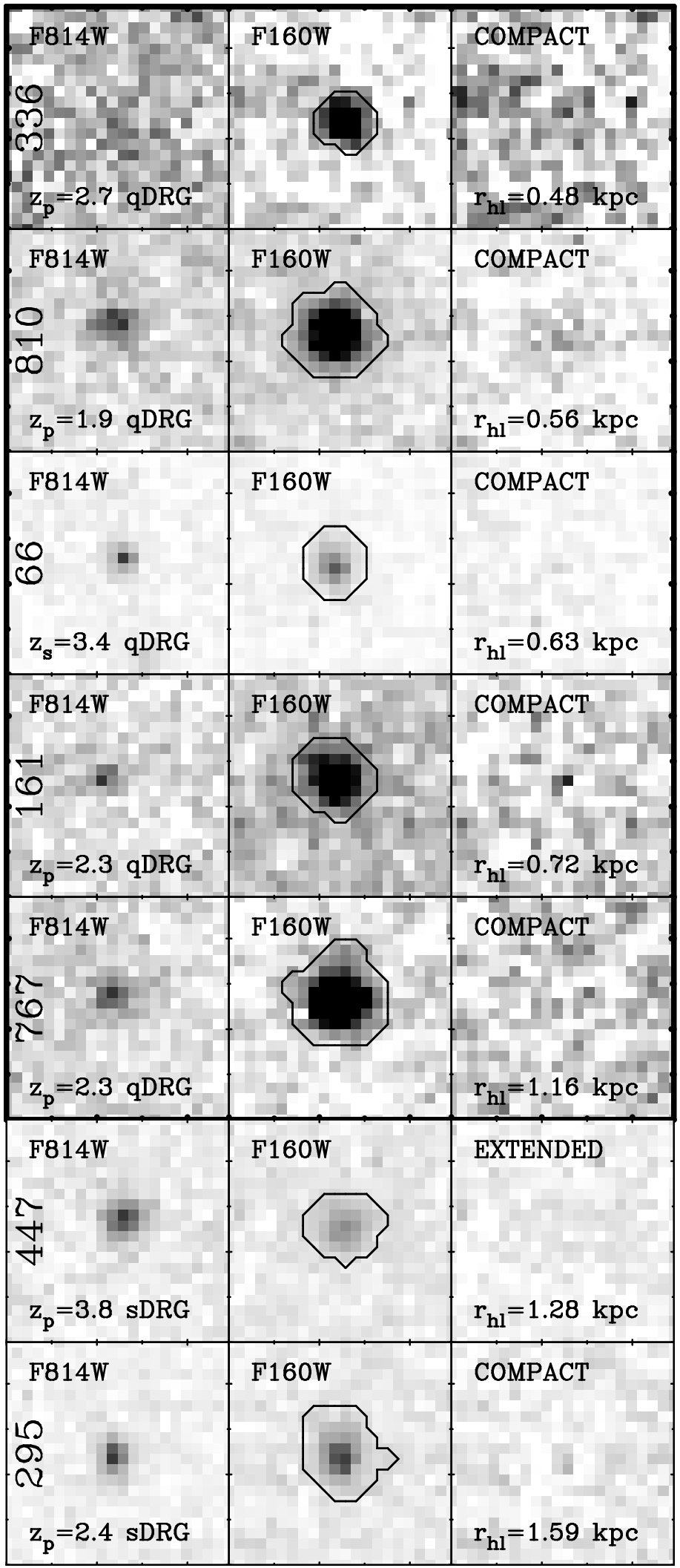}{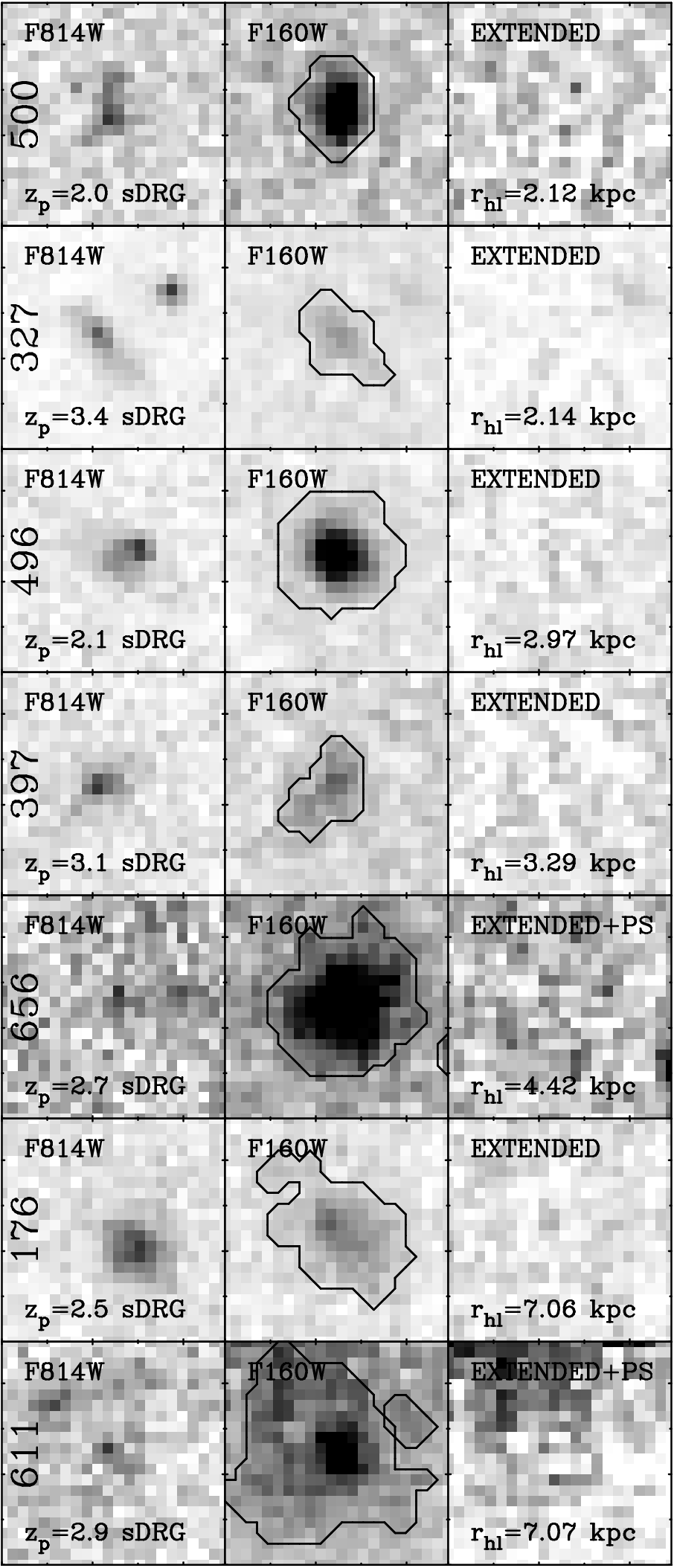}
\caption{WFPC2, NICMOS and model residual (left to right) image
cutouts of each of the 14 DRGs in the Hubble Deep Field South sorted
by galaxy size.  Each cutout has North up and East to the left and is
$2\farcs5$ on a side.  We have indicated the inferred star-formation
in the lower left corner of the WFPC2 cutouts, either `sDRG' for the
star-forming galaxies or `qDRG' for the quiescent sources.  The qDRG
cutouts are outlined with a thick black line.  The image contrasts
were chosen so that a galaxy with the same $F_{\lambda}$ would appear
the same.  The contour in the central image is the segmentation map
from object detection.  The contrast of the residual image is
different by a factor of 1.5 in $F_{\lambda}$ to show lower contrast
features.  The fits to the well-resolved galaxies 611 and 656 include
a central point source.
\label{fig:cuts}}
\end{figure}

We fit Sersic profiles \citep*{Sersic} to the data, allowing the shape
index ($n$) to take the values $n =$ 1 (exponential disk), 2, 3 and 4
($R^{1/4}$-law).  For each fit, we calculated the circularized
half-light radius ($r_{\rm hl} = \sqrt{ab}$) of the best-fitting
model.  The optical, NIR and model residual images for each galaxy are
shown in Figure~\ref{fig:cuts}.  We have defined a compact galaxy to
be one with an $r_e$ smaller than one pixel or resolved with a
best-fit $n=4$ profile.

The results are essentially unchanged whether the model or stellar PSF
was used for convolution.  The largest difference (but still within
the errors) was for the smallest source, 66, whose best-fit size
increased by 18\% when the stellar PSF was used.  For the other
galaxies the change was never larger than 15\% in either direction and
exhibited no systematic change.  The results from the two fitting
codes also agreed to within 5-10\% in every case with no observable
systematic trend.  Comparison between our galaxy sizes and those found
by Trujillo et al (2006) agree within a few percent for the two
galaxies which are well-resolved ($r_e > 0\farcs5$) in their
ground-based data.  For the compact galaxies the sizes derived using
the NICMOS data are systematically smaller by about 10\% than the
ground-based determinations when using the same Sersic $n$ values.

As with any galaxy fitting, our results may be skewed by deviations
from the assumed analytic profiles or by limitations of the data
themselves.  If an unresolved central source (active nucleus or
starburst) is present, the true profile will not be well-represented
by one of our analytic galaxy profiles.  Object 66 contains a
spectroscopically confirmed active nucleus, and object 767 shows an
excess at observed 8$\mu$m which may be due to an AGN.  However, the
strong break in the rest-frame optical and the high quality-of-fit for
these quiescent SEDs (see \S 4) argues against AGN significantly
biasing our size measurements based on the NICMOS data.  Furthermore,
none of the DRGs are detected in the radio imaging in the HDF-S
\citep*{Huynhetal05}.  Object 767 is detected in the {\it
Spitzer}/MIPS $24\mu$m imaging perhaps due to an AGN.  Our galaxy size
measurements could be biased to smaller values if as little as 10\% of
the light is coming from the nucleus \citep*[e.g.,][]{Daddietal05}.
However, even for extremely powerful radio galaxies at $z \sim 1$
essentially zero rest-frame UV light escapes the central region
\citep*{ZirmDickinsonDey03}.  Daddi et al. (2005) find several small,
quiescent galaxies at $z \sim 1.8$, two of which are very obscured
X-ray sources \nocite{Daddietal05}.  It is unclear how heavy
obscuration at X-ray wavelengths would not correspond to complete
nuclear extinction at rest-frame UV wavelengths.  Unfortunately, there
is no suitable X-ray data for this field.  We cannot rule out that
some of these DRGs contain powerful AGN but consider it unlikely that
AGN are responsible for every small derived galaxy size in this
sample.

\begin{figure}[b]
\epsscale{0.8}
\plotone{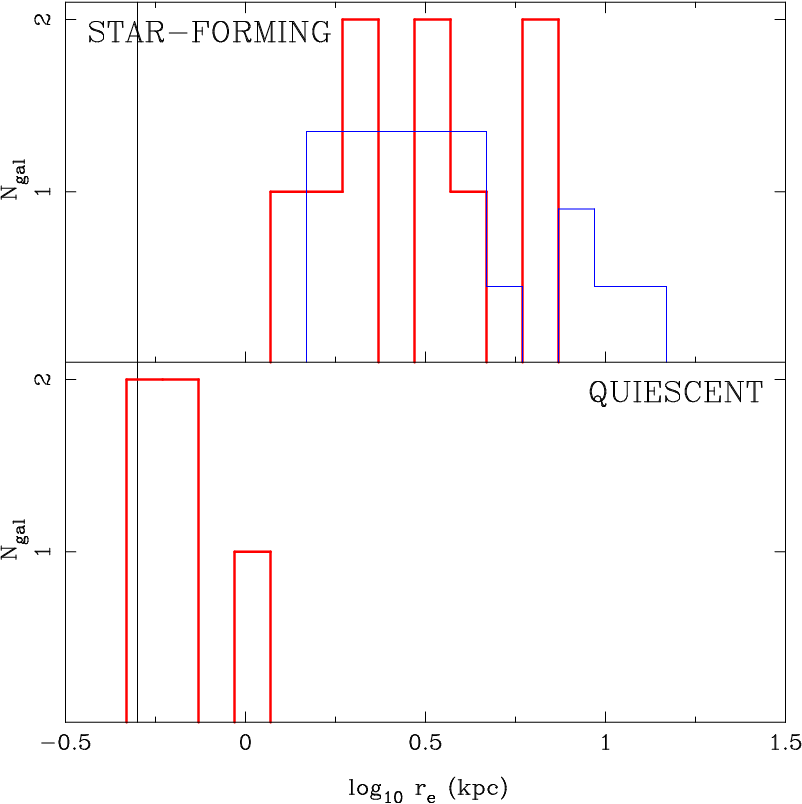}
\caption{Half-light radii distributions for the star-forming (upper
panel) and quiescent (lower panel) distant red galaxies (red) and
Lyman-break galaxies (blue).  The vertical line shows the average
physical resolution of the data.  The amount of overlap between the
quiescent and star-forming samples is minimal in galaxy size while
remaining substantial in the other observed properties (see
Fig.~\ref{fig:otherhist}).  This strongly suggests that a direct
correspondence exists between star formation and galaxy size for the
DRGs, rather than a mutual correlation with a third parameter.
\label{fig:sizes}}
\end{figure}

The surface-brightness (SB) limit for our imaging is $\sim 24.5$ AB
magnitudes arcsec$^{-2}$.  Lower SB features such as extended disks
may be undetected in our data and hence will be missing from our fits.
To search for low SB emission around the compact DRGs we stacked the
images of the compact galaxies to form a composite image.  This deeper
summed image shows no discernible evidence for an extended component.
Comparison of ground-based $H$-band and NICMOS total galaxy magnitudes
shows good agreement and no systematic offset indicating missed light
in the space-based imaging.  Furthermore, there is no obvious visual
indication of extended structures in the WFPC2 (rest-frame UV) data
for the compact galaxies.  We present the full size distributions in
the bottom panel of Figure~\ref{fig:sizes} and discuss their
implications in the next section.  The sizes are given in
Table~\ref{tab:drgs}.

As a comparison sample, we also fit profiles to all the galaxies in
the HDF-S which have been selected using the `Lyman-break' technique
(LBGs; Madau et al. 1996\nocite{Madauetal96}).  These galaxies were
selected and their photometric redshifts, masses and sizes derived
from the same exact dataset used for the DRGs.  They have the same
mean redshift as the DRG sample ($z=2.6$).  The size distribution of
the LBGs is also shown in Figure~\ref{fig:sizes}.

\section{Correlation Between Galaxy Size and Star Formation Rate}

\begin{figure}[t]
\epsscale{1.0}
\plotone{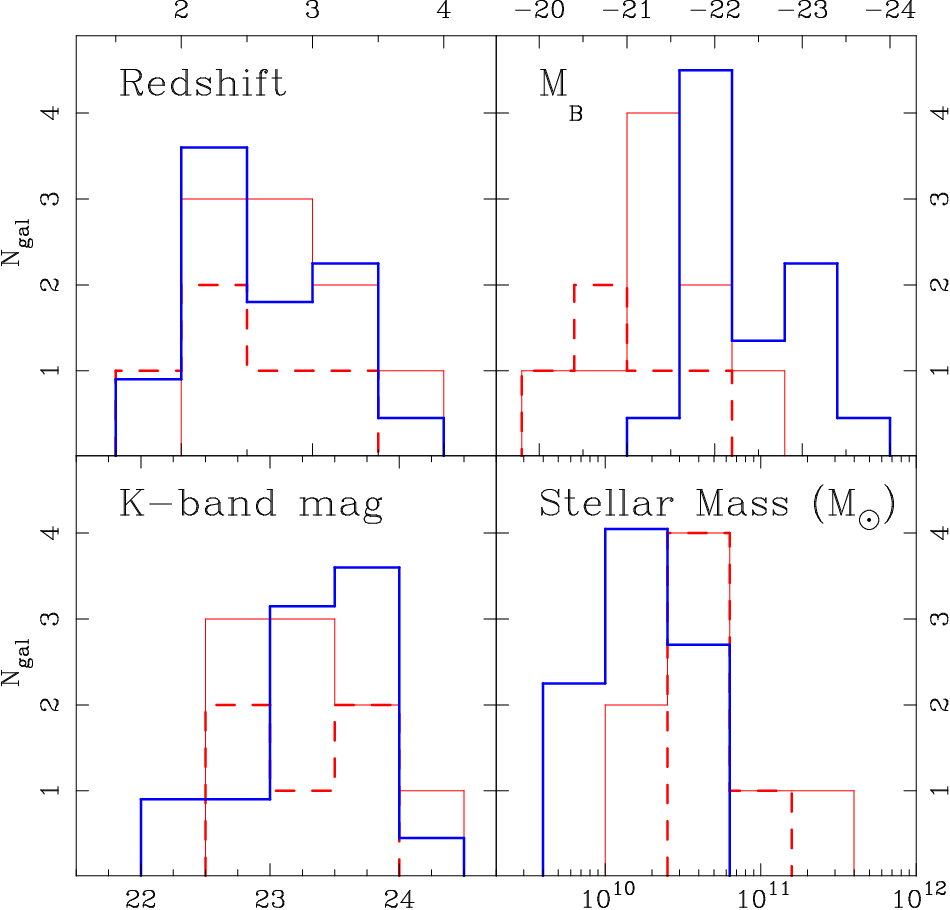}
\caption{Clockwise from Top Left: Redshift, luminosity, stellar mass
and apparent magnitude distributions for the star-forming (solid red
lines) and passive (dashed red) DRG and LBG (solid blue) populations.
The amount of overlap between the two samples is substantial in these
observed properties.  This strongly suggests that a direct
correspondence exists between star formation and galaxy size for the
DRGs, rather than a mutual correlation with a third parameter.
\label{fig:otherhist}}
\end{figure}

In Labb{\'e} et al. (2005) and Wuyts et al (2006, in
press\nocite{Wuytsetal06}), we have previously fit model spectral
energy distributions (SEDs) to the observed broad-band colors of the
DRGs to infer their luminosity-weighted stellar ages and derive
stellar masses.  The ground-based near-infrared and space-based
optical photometry were combined with data from {\it Spitzer}/IRAC to
construct the galaxies' SEDs.  In order to characterize the star
formation properties of the galaxies, we fitted two simple models from
Maraston (2005) to each source SED: a single stellar population
without dust, and a constant star formation model with dust.  We refer
to the galaxies best fitted by the dust-free SSP model as ``qDRGs'',
for ``quiescent'', and to the galaxies best fitted by the constant
star formation model as ``sDRGs'', for ``star-forming.''  We stress
that these models are simplifications, as in reality there is probably
a large range in star formation rates and ages.  This characterization
allows us to compare the properties of the galaxies with the highest
and lowest specific star formation rates.  These same data have been
used to derive stellar masses for the DRGs (see Table~\ref{tab:drgs};
Wuyts et al. 2006, in press\nocite{Wuytsetal06}).  As we will discuss
further below, we have decided to use Maraston (2005) models to derive
conservatively low masses for these galaxies.  The lack of
spectroscopic redshifts and emission line diagnostics precludes a more
detailed analysis \citep*[e.g.,][]{Krieketal06b}.  The dusty
star-forming population outnumbers the qDRGs by 9 to 5 in the HDF-S.

When we combine these results on the ages with the galaxy size
determinations from the NICMOS imaging we identify a clear trend for
the quiescent galaxies to be smaller than their star-forming cousins
at high redshift, with little overlap between the two galaxy types.
Figure~\ref{fig:sizes} show the galaxy effective radius distributions
for the two populations.  It is important to identify whether the
correlation involves a third parameter with which both morphology and
star-formation rate are individually related, e.g. redshift,
luminosity or stellar mass.  In Figures~\ref{fig:sizes} and
\ref{fig:otherhist} we show the size, redshift, luminosity, mass and
apparent magnitude distributions for the two sub-populations and the
LBG comparison sample.  It is clear that the overlap between the two
DRG sub-samples is greater in redshift, mass, luminosity and apparent
magnitude than in size.  Therefore it appears that there is a direct
correlation between galaxy size and stellar age.  Additionally,
comparison with the sizes of other star-forming galaxies (LBGs) in the
same redshift range shows that their size distribution is consistent
with that of the star-forming DRGs (see Figure~\ref{fig:sizes}).

This trend of older stellar populations to be contained in compact
structures is qualitatively consistent with the morphology-color
correlation observed in the local universe
\citep*{Kauffmannetal03b,Brinchmannetal04}.  More generally, the
existence of compact galaxies with little or no star-formation
supports the suggestion by Faber et al. (2005) and others that the
same event is responsible for both the cessation of star-formation and
the morphological transformation from late- to early-type.
Furthermore, our results show that massive, centrally-concentrated
galaxies exist within $\approx 2.5$ Gyrs of the Big Bang.

\section{qDRG Mass Densities}

\begin{figure*}
\epsscale{0.9}
\plotone{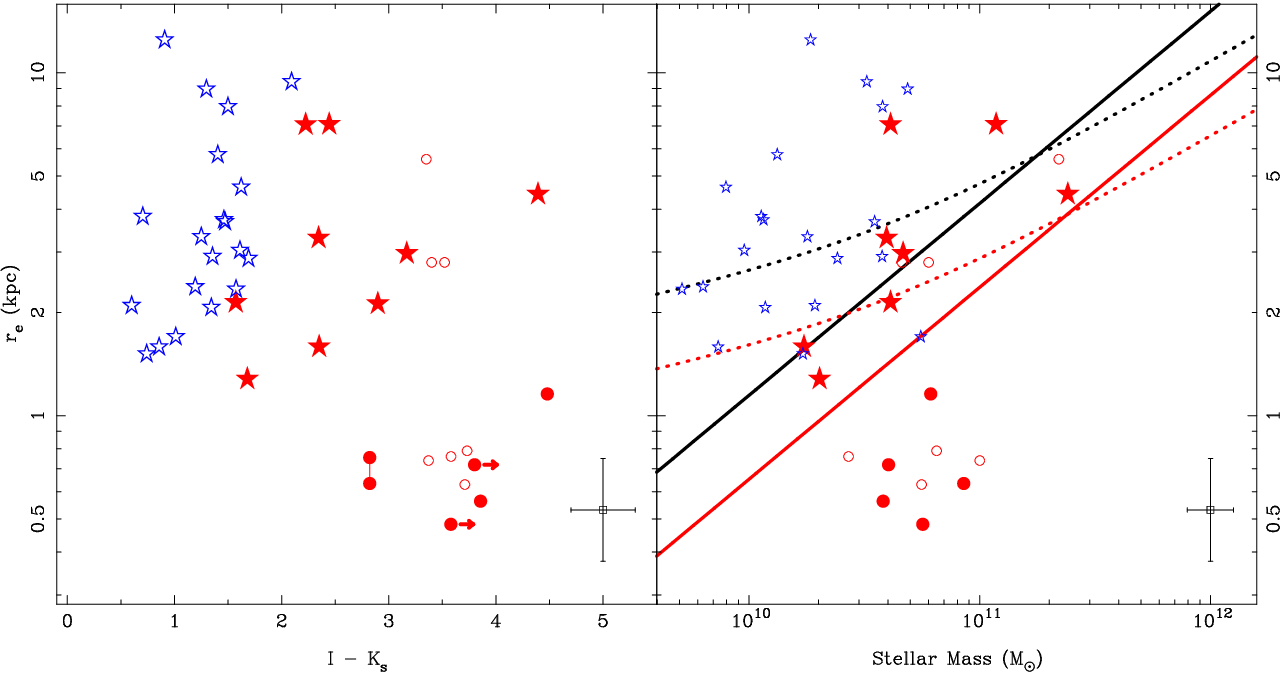}
\caption{Left panel: Half-light radius vs observed $I-K_{s}$ color for
the star-forming (red stars) and quiescent (filled red circles) DRGs and
star-forming LBGs (blue stars) in the HDF-S.  The open red circles
are the quiescent $z \sim 1.8$ galaxies from Daddi et
al. (2005)
Two of the passive sources are
undetected in I and are therefore shown as lower-limits (arrows).  The
error bar in the lower right is representative of the fits to the
quiescent galaxies.  The filled red circles connected by a line show the
maximal offset introduced by using a stellar rather than a model PSF
(an increase in size for object 66 of 18\%).  Right panel: Half-light
radius vs. stellar mass with the same symbols.  The two sets of
over-plotted lines are the size-mass relationships derived by Shen et
al. 2003 for early-type (black solid) and late-type (black dotted)
galaxies in the local universe and redshifted to $z=2.5$ using the
inferred size-redshift evolution for $M_{\star} > 3 \times 10^{10}
M_{\odot}$ galaxies from Trujillo et al. (2006; red lines).
\label{fig:sizeIK}
}
\end{figure*}

The quiescent distant red galaxies (qDRGs) are of particular interest
because of their small sizes, high inferred stellar masses and implied
early formation.  Their large mass of stars and lack of vigorous
current star-formation suggest that they have had at least one major
star-formation episode which has now ceased.  In the left panel of
Figure~\ref{fig:sizeIK} we plot the half-light radius versus the
observed $I-K_{s}$ color.  For $z>2$ galaxies, redder $I-K_{s}$ colors
can either indicate lower specific star formation rates or higher dust
content \citep*{Labbeetal05}.  The quiescent DRGs (filled red circles)
are clearly separated from the star-forming populations in this plot.
The two red circles connected by a line indicates the maximal offset
due to convolution with different PSFs for object 66.

In the right panel of Figure~\ref{fig:sizeIK} we show half-light
radius versus stellar mass for the DRGs and LBGs in the HDF-S.  The
over-plotted lines show the size-mass relation for local early-type
(solid line) and late-type (dotted line) galaxies ($z < 0.1$) from the
Sloan Digital Sky Survey (SDSS; Shen et al. 2003\nocite{Shenetal03}).
It is clear that the qDRGs are rather different from local
early-types.  Recent studies of the evolution in the size-mass
relation \citep*{Nachoetal04,Nachoetal06} find that galaxies with
comparable stellar masses to the qDRGs were a factor 1.7 smaller at
higher redshifts than locally, qualitatively consistent with the
observed trend in our data.  However the qDRGs are still much smaller
than the evolution would predict.  This is likely due to the inclusion
of the full galaxy population to derive the evolution and scatter in
the evolution prediction.  Our results also agree with direct
(rest-frame UV) size measurements of quiescent galaxies at $z \sim
1.8$ \citep*{Daddietal05}.  However, by $z \sim 1$ the early-type
population does not show many galaxies in the size-mass region of the
qDRGs \citep*{McIntoshetal05}.

\begin{figure*}[t]
\plotone{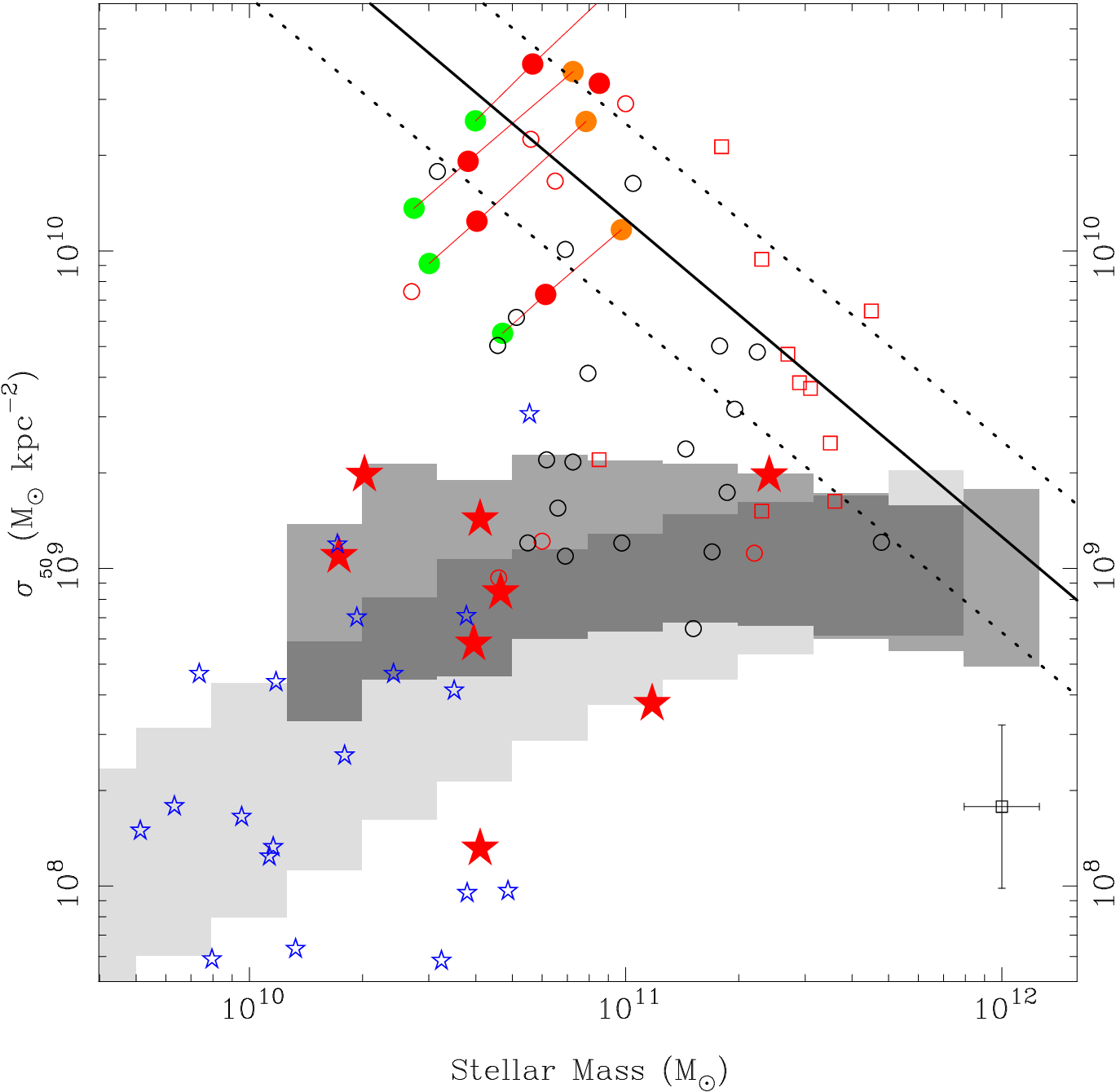}
\caption{Average surface mass density within the half-light radius.
The sDRGs (red stars) and qDRGs (filled red circles) are shown using the
Maraston \citep*{Maraston05} SED stellar mass estimates from Wuyts et
al. (2006, in press).  The filled green and orange circles 
illustrate the uncertainty due to photometric redshift errors 
for the 4 qDRGs without spectroscopic redshifts.  
The local values for early-type (dark grey
shading) and late-type (light grey shading) galaxies
\citep*{Shenetal03} are over-plotted.  The solid line shows the simple
trend of surface density with mass expected for dissipationless
mergers normalized to the qDRGs
\citep*[e.g.,][]{NipotiLondrilloCiotti03}; the dotted lines show the
trend if our mass estimates are systematically too high or low by a
factor of two.  The open black circles are $z\sim1$ ellipticals
\citep*{vanderWeletal06}.  The open red circles are passive galaxies
found by Daddi et al. (2005)
in the UDF at $z\sim 1.5$ using their (lower) Maraston SED mass
estimates \citep*{Marastonetal06} and the open red squares are dense,
passive sources found in the MUNICS survey \citep*{Trujilloetal06b}.
The blue stars are Lyman-break galaxies from this work.  A
characteristic error bar for the qDRGs is shown as an open square in
the lower right-hand corner.
\label{fig:density}
}
\end{figure*}

Perhaps most striking is the difference shown in
Figure~\ref{fig:density} where we plot the average surface stellar
mass density within the half-light radius, $\sigma_{50}$, versus the
stellar mass.
\begin{equation}
\sigma_{50} = \frac{0.5 M_{\star}}{\pi r_{e}^{2}} 
\end{equation}
The sDRGs and LBGs overlap the region of the local galaxy samples.
The much higher densities of the qDRGs suggest that substantial
downward density evolution must take place between $z \sim 2$ and the
present-day.  However, it is nearly impossible to lower the density of
a stellar distribution via secular evolution on this timescale.  The
relaxation time for these compact DRGs is still much longer than the
Hubble time, despite their high densities.  Dissipationless, or `dry',
merging \citep*[e.g.,][]{vanDokkum05} would predict a linear decrease
in the surface density with accumulated mass
\citep*[e.g.,][]{NipotiLondrilloCiotti03}.  The diagonal lines in
Fig.~\ref{fig:density} show this power-law trend with a \emph{fiducial
normalization to the qDRGs}.  These lines also happen to pass through
some of the $z \sim 1$ galaxies which are also ``overdense'' compared
to local early-types.  These $z \sim 1$ sources are clearly
ellipticals (i.e., they lie on the Fundamental Plane and follow
$r^{1/4}$-law profiles) and these are their stellar masses from
comparable SED fitting, also using a Salpeter IMF
\citep*{vanderWeletal05,vanderWeletal06}.  

Alternatively, the qDRGs may be something of a mixed population; both
in the sense that the color and SED selection collects multiple galaxy
types and that measurements may lead to misclassification of some
sources.  Errors in the photometric redshift determination and SED
modeling may still allow significant uncertainty in the ``compact,
evolved stellar population'' interpretation (see Table~\ref{tab:drgs}
for the photo-$z$ errors).  Of the five qDRGs only one has a
spectroscopic redshift, object 66, and this source harbors an active
nucleus.  Also plotted in Fig.~\ref{fig:density} are the offsets in
mass and density due to the photometric redshift errors.  The
1$\sigma$ errors from the phot-$z$ fits are used to estimate the
corresponding errors on the stellar mass and density.  The filled
green circles are the low redshift estimates and the filled orange
circles are the high redshift estimate.  These errors do not account
for a change in the preferred SED template (and therefore a change in
the assumed mass-to-light ratio) due to the change in redshift.  Using
different stellar population synthesis models to fit the SEDs result
in significant differences in the derived properties.  In particular,
the Maraston (2005) models \nocite{Maraston05} derive younger ages and
consequently smaller, more conservative, stellar masses for the qDRGs
on average and we use these in Figures~\ref{fig:otherhist},
\ref{fig:sizeIK} and \ref{fig:density}.  This offset is reflected by
the dotted lines in of Fig.~\ref{fig:density}.  A similar offset would
result by allowing for dust in the quiescent SED models.
Overestimation of the galaxy masses could also be due to a different
stellar IMF.  We note, for example, that Baugh et
al. (2005)\nocite{Baughetal05} required a top-heavy IMF to model the
sub-mm source counts correctly.  Regardless of this particular
result's merit, it is certainly clear that a non-constant IMF renders
the mass estimates very uncertain.  Direct mass estimates are
therefore extremely important (but remain very difficult with current
facilities).  It should also be noted that the optical-NIR colors of
the qDRGs exclude the possibility that they are stars
\citep*{Franxetal03}.

\section{Discussion and Conclusions}

We have presented high spatial-resolution rest-frame optical imaging
data for NIR-selected Distant Red Galaxies in the HDF-S.  The
combination of these data with our modeling of their broad-band
optical-infrared SEDs has highlighted the `quiescent' population,the
qDRGs, as being particularly dense, massive stellar cores at
high-redshift.  The evolution of the qDRGs into their counterparts at
low redshift is problematic.  Their relatively high stellar masses,
the strong clustering \citep*{Daddietal03,Quadrietal06} of the DRGs as
a population and the small sizes of the qDRGs in particular suggest
that qDRGs are destined to become massive concentrated galaxies at
lower redshift, i.e., ellipticals.  However, their surface stellar
mass densities are more than an order of magnitude higher than local
ellipticals.  Mergers which involve substantial gas dissipation may
increase the mass density of the remnant \citep*{Robertsonetal06},
while dissipationless, or `dry', merging increases the size of the
galaxies sufficiently to lower their densities \citep*[e.g.,
][]{NipotiLondrilloCiotti03,vanDokkum05}.  If the qDRGs were to evolve
via dry merging they would only reach local galaxy densities at very
high masses ($\simgt 10^{12} M_{\odot}$).  It is unlikely that the
small HDF-S field would contain five progenitors of such rare, giant
galaxies.  However, simulations do show that the progenitors of giant
local galaxies may be spread over an area comparable to the HDF-S at
high redshift \citep*[e.g.,][]{Gaoetal04}. If these qDRGs were to
evolve along the dry merging line, they would require more than 4
equal mass mergers to match the density of the $z \sim 0$ ellipticals.
So while it may seem numerically possible that these qDRGs would merge
with each other to produce a single giant galaxy, and the uncertain
photometric redshifts in this field cannot definitively exclude this
possibility, it would imply that the HDF-S is a rather special (rare)
field.  They may also merge with the more numerous star-forming
galaxies as long as the remnant is less dense than the compact DRG.
Interestingly, Robertson et al. (2006) find that the role of
dissipation decreases as the mass of the merging galaxies increase.
More and more of the gas is dynamically heated into the galactic halo
and is unable to cool.  It is worth noting that despite the
possibility of misclassification of the qDRGs as compact, even a
single dense stellar core at high-redshift would require a plausible
evolutionary scenario to match local galaxy properties.

We also want to know how a dense galaxy may have formed.  From the
modeling of their SEDs, the qDRGs have inferred luminosity-weighted
stellar ages using the Maraston (Bruzual \& Charlot) models between
0.3 (0.5) and 2.0 (2.3) Gyr (Wuyts et al. 2006\nocite{Wuytsetal06}).  Mergers
which involve substantial gas dissipation seem to result in high mass
density of the remnants \citep*{Robertsonetal06}.  Such gas-rich
mergers are more likely at higher redshift where relatively little of
the gaseous content of galaxies has already been used in
star-formation \citep*{KhochfarSilk06a,
KhochfarSilk06b,NaabKhochfarBurkert06} and the generic merger rate is
much higher.  Further data on these and similar galaxies at higher
redshift will help distinguish the origin of these dense galactic
cores.

Near-infrared spectroscopy can confirm the redshifts and help
determine whether the light is indeed dominated by stars rather than
AGN \citep*{Krieketal06b} and may eventually enable kinematic mass
estimates.  Initial results from a NICMOS study of a wider field
confirms the trend toward small sizes and high densities for the qDRGs
quoted here (Toft et al., in preparation).  As samples of qDRGs
continue to grow, further high spatial-resolution imaging (either from
NICMOS/WFC3 or from adaptive-optics systems on the ground) will be
able to discover whether this trend toward small sizes observed in the
HDF-S persists over larger volumes and lower galaxy masses and
determine the extent to which AGN contribute to the phenomenon.

\acknowledgments 

Support for program \# 9723 was provided by NASA through a grant
(GO-09723.01-A) from the Space Telescope Science Institute, which is
operated by the Association of Universities for Research in Astronomy,
Inc., under NASA contract NAS 5-26555.  The entire FIRES team thanks
the staff of the Lorentz Center in Leiden for their hospitality and
excellent meeting facilities.  AWZ gratefully acknowledges funding
from NWO during his time in Leiden and thanks Colin Norman for helpful
discussions.  We thank the anonymous referee for useful suggestions
and their prompt attention to the manuscript.


\begin{deluxetable}{rcrccccc}
\tablecolumns{8}
\tablewidth{0pc}
\tablecaption{Properties of the HDF-S DRGs\label{tab:drgs}}
\tablehead{
\colhead{ID}  &  \colhead{$z_{\rm phot}$\tablenotemark{a}} & \colhead{SED Type\tablenotemark{b}} & \colhead{$I-K_{s}$} & \colhead{$r_e$} & \colhead{$r_e$} & \colhead{Stellar Mass} & \colhead{$\sigma_{50}$\tablenotemark{c}} \\
\colhead{} & \colhead{} & \colhead{} & \colhead{(AB)} & \colhead{(arcsec)} & \colhead{(kpc)} & \colhead{($10^{11}$ $M_{\odot}$)} & \colhead{($10^{10}$ $M_{\odot}$ kpc$^{-2}$)}}
\startdata

336 & $2.7_{-0.1}^{+6.7}$ & quiescent & 3.6 & 0.06 & 0.48 & $0.6_{-0.2}^{+391}$\tablenotemark{e} & 3.88 \\ 
810 & $1.9_{-0.1}^{+0.1}$ & quiescent & 3.9 & 0.07 & 0.56 & $0.4_{-0.1}^{+0.3}$\tablenotemark{e} & 1.92 \\ 
66 & 3.4\tablenotemark{d} & quiescent & 2.8 & 0.08 & 0.63 & 0.9 & 3.36 \\ 
161 & $2.3_{-0.1}^{+0.3}$ & quiescent & 3.8 & 0.09 & 0.72 & $0.4_{-0.1}^{+0.4}$\tablenotemark{e} & 1.25 \\ 
767 & $2.3_{-0.1}^{+0.1}$ & quiescent & 4.5 & 0.14 & 1.16 & $0.6_{-0.1}^{+0.4}$\tablenotemark{e} & 0.73 \\ 
447 & $3.8_{-0.2}^{+0.3}$ & star-forming & 1.7 & 0.18 & 1.28 & 0.2 & 0.20 \\ 
295 & $2.4_{-0.1}^{+0.6}$ & star-forming & 2.4 & 0.19 & 1.59 & 0.2 & 0.11 \\ 
500 & $2.0_{-0.1}^{+0.3}$ & star-forming & 2.9 & 0.25 & 2.12 & 0.3 & 0.09 \\ 
327 & $3.4_{-0.2}^{+0.2}$ & star-forming & 1.6 & 0.28 & 2.14 & \nodata\tablenotemark{f} & \nodata\tablenotemark{f} \\ 
496 & $2.1_{-0.2}^{+0.1}$ & star-forming & 3.2 & 0.35 & 2.97 & 0.5 & 0.08 \\ 
397 & $3.1_{-0.8}^{+0.1}$ & star-forming & 2.3 & 0.42 & 3.29 & 0.4 & 0.06 \\ 
656 & $2.7_{-0.2}^{+0.3}$ & star-forming & 4.4 & 0.55 & 4.42 & 2.4 & 0.20 \\ 
176 & $2.5_{-0.1}^{+0.3}$ & star-forming & 2.2 & 0.86 & 7.06 & 0.4 & 0.01 \\ 
611 & $2.9_{-0.1}^{+0.1}$ & star-forming & 2.4 & 0.89 & 7.07 & 1.2 & 0.04 \\ 
\enddata
\tablenotetext{a}{Photometric redshifts as derived in Rudnick et al. 2001, 2003}
\tablenotetext{b}{SED Type derived from Maraston (2005) population synthesis models in Wuyts et al. (2006, in press)}
\tablenotetext{c}{Average stellar surface mass density within the effective radius}
\tablenotetext{d}{This object, 66, has a spectroscopic redshift}
\tablenotetext{e}{Errors on the masses of the qDRGs due to the errors on the photometric redshifts}
\tablenotetext{f}{Object 327 is confused in the $K_{S}$-band data and is therefore excluded from the stellar mass analysis}
\end{deluxetable}

\end{document}